\documentclass[aps,showpacs,preprintnumbers,amsmath,amssymb]{revtex4}
\usepackage{dcolumn}
\usepackage[dvips]{epsfig}

\begin{document}
\baselineskip=0.8 cm
\title{\bf Thin accretion disk around a rotating Kerr-like black hole in Einstein-bumblebee gravity model}
\author{ Changqing Liu$^1$}\email{lcqliu2562@163.com}
\author{Chikun Ding$^{1,2}$}\email{Chikun_Ding@huhst.edu.cn}
\author{Jiliang Jing$^{2}$}\email{jljing@hunnu.edu.cn}

  \affiliation{$^1$Department of Physics, Hunan University of Humanities, Science and Technology, Loudi, Hunan
417000, P. R. China\\
$^2$Key Laboratory of Low Dimensional
Quantum Structures and Quantum Control of Ministry of Education,
and Synergetic Innovation Center for Quantum Effects and Applications,
Hunan Normal University, Changsha, Hunan 410081, P. R. China}

\begin{abstract}
\baselineskip=0.6 cm
\begin{center}
{\bf Abstract}
\end{center}

We study the accretion process in the thin disk around a rotating
Kerr-like black hole in Einstein-bumblebee gravity model where Lorentz
symmetry is spontaneously broken once a vector field acquires a vacuum
expectation value. In the present paper we obtain the energy flux, the emission spectrum and accretion efficiency from the
accretion disks around the rotating Kerr-like black hole, and we compare
them to the general Kerr case. These significant features
in the mass accretion process may provide a possibility to test whether the Lorentz
symmetry is spontaneously broken or not in the Einstein-bumblebee gravity model by future astronomical
observations.

\end{abstract}

 \maketitle
\newpage
\section{Introduction}

General relativity(GR) describes gravitation at a classical level. Standard Model (SM) describes particles and
the other three fundamental interactions at a quantum
level. In the past decades, there were some attempts to construct a fundamental theory that unifies between  GR and SM theories.
The unification between these two theories imply that small Lorentz violation(LV)  \cite{PhysRevD.39.683,Kost1991} effects may appear at
 the Planck scale and Lorentz symmetry breaking
arises in the context of string theory \cite{PhysRevD.39.683}, noncommutative field
theories \cite{Carroll} or loop quantum gravity theory \cite{Gambini}. Some novel hypothetical
effects that break local Lorentz symmetry and CPT symmetry in gravitational
experiments as well as solar system and astrophysical observations
have been studied in recent works. Much of this work uses the effective
field theory framework. Thus, the study of Lorentz violation is a valuable tool to probe
the foundations of modern physics. These studies include LV in the neutrino sector \cite{dai},
the standard-model extension (SME) \cite{colladay}, LV in the non-gravity sector \cite{coleman},
and LV effect on the formation of atmospheric showers \cite{rubtsov}.

The SME is an effective field theory describing the SM coupled to GR, allowing for dynamical curvature modes, and includes additional terms containing information about the LV occurring at the Plank scale. The LV terms in the SME take the form of Lorentz-violating operators coupled to coefficients with Lorentz indices. The presence of LV in a local Lorentz frame is signaled by a nonzero vacuum value for one or more quantities carrying local Lorentz indices. The so-called bumblebee model, as one kind of the SME in the effective field theory, was first used by Kostelecky and Samuel in 1989 \cite{PhysRevD.39.683}. It is a simple model for investigating the consequences
of spontaneous LV. The LV arises from the dynamics of a single vector or axial-vector field $B_\mu$, known as the bumblebee field. It is ruled by a potential exhibiting a minimum rolls to its vacuum expectation value. Several applications with the bumblebee
model have been done, such as, traversable wormhole solution in the framework of the bumblebee
gravity theory \cite{Ovgun:2018xys}, exact Schwarzschild-like solution\cite{Casana:2017jkc}, cosmological implications of Bumblebee
vector models \cite{Capelo:2015ipa,bs}, G\"{o}del solution \cite{Nascimento:2014vva}, quantum effects \cite{Her,Maluf:2015hda}. Recently, we obtain an exact Kerr-like black hole solution \cite{ding} by solving the corresponding gravitational field equations in Einstein-bumblebee gravity model where Lorentz symmetry is spontaneously broken once a vector field acquires a vacuum expectation value.

Accretion disks are well known observationally, representing flattened
astronomical structures. The accretion processes is a powerful
indicator of the physical nature of the central celestial objects,
which means that the analysis of the signatures of the accretion
disk around the rotating Kerr like black hole could help us to detect
the Lorentz breaking effects in Einstein-bumblebee gravity model. The steady-state thin accretion disk model is the
simplest theoretical model of the accretion disks, in which the disk
has negligible thickness so that the heat generated by stress and
dynamic friction in the disk can be dispersed through the radiation
over its surface \cite{sdk1,sdk2,Page,Thorne,Luminet}. This cooling
mechanism ensures that the disk can be in hydrodynamical equilibrium
and the mass accretion rate in the disk maintains a constant, which
is independent of time variable. The physical properties of matter
forming a thin accretion disk in a variety of background spacetimes
have been investigated extensively in
\cite{JGa,Harko,Bhattacharyya,Kovacs,Torres,Yuan,Guzman,Pun,Cs,ZLP,CBa3,CBa4,Bhattacharyya1}. The special signatures appeared in the energy flux and the emission
spectrum emitted by the disk can provide us not only the information
about black holes in the Universe, but also the profound
verification of alternative theories of gravity.
 Therefore, in this paper, we shall focus on the influence of the Lorentz breaking parameter on the properties of the thin accretion disk around a Kerr like black hole.

The rest of the paper is organized as follows: in section II, we will
review briefly the rotating Kerr like black hole metric in Einstein-bumblebee gravity model, and then present the geodesic equations for
the timelike particles moving in the equatorial plane in this
background. In Sec.III, we study the physical properties of the thin
accretion disk around the rotating Kerr like black hole  and probe the
effects of the Lorentz breaking parameter on the energy flux, temperature
and emission spectrum of the thin accretion disks onto this black
hole. Sec. IV is devoted to a summary.

\section{The geodesic equations in the rotating Kerr-like black hole}

Let us now review briefly the exact Kerr-like black hole  \cite{ding}
In the bumblebee gravity theory, the bumblebee vector field $B_{\mu}$ acquires a nonzero vacuum expectation value, under a suitable potential, inducing a spontaneous Lorentz symmetry breaking in the gravitational sector. It is
 described by the  action,
\begin{eqnarray}
\mathcal{S}=
\int d^4x\sqrt{-g}\Big[\frac{1}{16\pi G_N}(\mathcal{R}+\varrho B^{\mu}B^{\nu}\mathcal{R}_{\mu\nu})-\frac{1}{4}B^{\mu\nu}B_{\mu\nu}
-V(B^{\mu}B_{\mu}\pm b^2)\Big], \label{action}
\end{eqnarray}
where $\varrho$ is a real coupling constant (with mass dimension $-1$) which controls the non-minimal gravity interaction to bumblebee field $B_\mu$ (with the mass dimension 1), and $b^2$ is a real positive constant. The bumblebee field strength is defined by
\begin{eqnarray}
B_{\mu\nu}=\partial_{\mu}B_{\nu}-\partial_{\nu}B_{\mu}.
\end{eqnarray}
The potential $V$, driving Lorentz and/or $CPT$ (charge, parity and time), have a minimum at $B^{\mu}B_{\mu}\pm b^2=0$ and $V'(b_{\mu}b^{\mu})=0$, where $b^{\mu}=\langle B^{\mu}\rangle$ and has constant magnitude $b_{\mu}b^{\mu}=\mp b^2$.

The action (\ref{action}) yields the gravitational field equation in vacuum
\begin{eqnarray}\label{einstein}
\mathcal{R}_{\mu\nu}=\kappa T_{\mu\nu}^B+2\kappa g_{\mu\nu}V
+\frac{1}{2}\kappa g_{\mu\nu} B^{\alpha\beta}B_{\alpha\beta}-
\kappa g_{\mu\nu} B^{\alpha}B_{\alpha}V'+\frac{\varrho}{4}g_{\mu\nu}\nabla^2(B^{\alpha}B_{\alpha})
+\frac{\varrho}{2}g_{\mu\nu}\nabla_{\alpha}\nabla_{\beta}(B^{\alpha}B^{\beta}),
\end{eqnarray}
where $\kappa=8\pi G_N$ and the bumblebee energy momentum tensor $T_{\mu\nu}^B$ is
\begin{eqnarray}
&&T_{\mu\nu}^B=-B_{\mu\alpha}B^{\alpha}_{\;\nu}-\frac{1}{4}g_{\mu\nu} B^{\alpha\beta}B_{\alpha\beta}- g_{\mu\nu}V+
2B_{\mu}B_{\nu}V'\nonumber\\
&&+\frac{\varrho}{\kappa}\Big[\frac{1}{2}g_{\mu\nu}B^{\alpha}B^{\beta}R_{\alpha\beta}
-B_{\mu}B^{\alpha}R_{\alpha\nu}-B_{\nu}B^{\alpha}R_{\alpha\mu}\nonumber\\
&&+\frac{1}{2}\nabla_{\alpha}\nabla_{\mu}(B^{\alpha}B_{\nu})
+\frac{1}{2}\nabla_{\alpha}\nabla_{\nu}(B^{\alpha}B_{\mu})
-\frac{1}{2}\nabla^2(B^{\mu}B_{\nu})-\frac{1}{2}
g_{\mu\nu}\nabla_{\alpha}\nabla_{\beta}(B^{\alpha}B^{\beta})\Big].
\end{eqnarray}

The bumblebee field is fixed to be
\begin{eqnarray}
B_\mu=b_\mu,
\end{eqnarray}
and $V=0,\;V'=0$. Then Eq. (\ref{einstein}) leads to gravitational field equations
\begin{eqnarray}\label{bar}
\bar R_{\mu\nu}=0,
\end{eqnarray}
with
\begin{eqnarray}
&&\bar R_{\mu\nu}=\mathcal{R}_{\mu\nu}+\kappa b_{\mu\alpha}b^{\alpha}_{\;\nu}-\frac{\kappa}{4}g_{\mu\nu} b^{\alpha\beta}b_{\alpha\beta}+\varrho b_{\mu}b^{\alpha}\mathcal{R}_{\alpha\nu}
+\varrho b_{\nu}b^{\alpha}\mathcal{R}_{\alpha\mu}
-\frac{\varrho}{2}g_{\mu\nu}b^{\alpha}b^{\beta}\mathcal{R}_{\alpha\beta}+\bar B_{\mu\nu},\nonumber\\
&&\bar B_{\mu\nu}=-\frac{\varrho}{2}\Big[
\nabla_{\alpha}\nabla_{\mu}(b^{\alpha}b_{\nu})
+\nabla_{\alpha}\nabla_{\nu}(b^{\alpha}b_{\mu})
-\nabla^2(b_{\mu}b_{\nu})\Big].
\end{eqnarray}
 In the standard
Boyer-Lindquist coordinates, the metric of this rotating Kerr-like
black hole has a form \cite{ding}
\begin{eqnarray}
ds^2=g_{tt}dt^2+g_{rr}dr^2+g_{\theta\theta}d\theta^2+g_{\phi\phi}
d\phi^2+2g_{t\phi}dtd\phi, \label{metric0}
\end{eqnarray}
where
\begin{eqnarray}
g_{tt}&=&- \Big(1-\frac{2Mr}{\rho^2}\Big),\;\;\;\;\;
g_{t\phi}=-\frac{2Mra\sqrt{1+\it l}\sin^2\theta}{\rho^2},\nonumber\\
g_{rr}&=&\frac{\rho^2}{\Delta},\;\;g_{\theta\theta}=\rho^2,\;\;g_{\phi\phi}=\frac{A\sin^2\theta}{\rho^2},
\end{eqnarray}
with
\begin{eqnarray}
\rho^2=r^2+(1+\it l)a^2\cos^2\theta,\;\;\;
\Delta=\frac{r^2-2Mr}{1+\it l}+a^2,\;\;A=\big[r^2+(1+\it l)a^2\big]^2-\Delta(1+\it l)^2 a^2\sin^2\theta.
\end{eqnarray}
Here $\it l$ is the LV parameter. As $\it l=0$, the black hole is
reduced to the usual Kerr black hole in general relativity. The metric (\ref{metric0}) represents a purely radial Lorentz-violating  black hole solution with rotating angular momentum $a$. It is singular at $\rho^2=0$ and at $\Delta=0$. The solution of $\rho^2=0$  is a ring shape physical
singularity at the equatorial plane of the center of rotating black hole with radius $a$. Its event horizons and ergosphere locate at
\begin{eqnarray}
r_{\pm}=M\pm\sqrt{M^2-a^2(1+\it l)},\;r^{ergo}_{\pm}=M\pm\sqrt{M^2-a^2(1+\it l)\cos^2\theta},
\end{eqnarray}
where $\pm$ signs correspond to outer and inner horizon/ergosphere, respectively. It is easy to see that there exists a black hole if and only if
\begin{eqnarray}
|a|\leq \frac{M}{\sqrt{1+\it l}}.
\end{eqnarray}

According to the thin accretion disk model, one can assumes that the
disk is on the equatorial plane and that the matter moves on nearly
geodesic circular orbits. In the rotating Kerr-like black hole
spacetime (\ref{metric0}), the time-like geodesics equations of a
particle can be expressed as
\begin{eqnarray}
&&u^{t}=\frac{dt}{d\lambda}=\frac{\tilde{E}g_{\phi\phi}-\tilde{L}g_{t\phi}}{g^2_{t\phi}-g_{tt}g_{\phi\phi}},\label{u1}\\
&&u^{\phi}=\frac{d\phi}{d\lambda}=\frac{\tilde{E}g_{t\phi}+\tilde{L}g_{tt}}{g^2_{t\phi}-g_{tt}g_{\phi\phi}},\label{u2}\\
&&g_{rr}\bigg(\frac{dr}{d\lambda}\bigg)^2+g_{\theta\theta}\bigg(\frac{d\theta}{d\lambda}\bigg)^2=V_{eff},
\end{eqnarray}
with the effective potential
\begin{eqnarray}
V_{eff}=\frac{\tilde{E}^2g_{\phi\phi}+2\tilde{E}\tilde{L}g_{t\phi}+\tilde{L}^2g_{tt}}{g^2_{t\phi}-g_{tt}g_{\phi\phi}}-1,
\end{eqnarray}
where $\tilde{E}$ and $\tilde{L}$ are the specific energy and the
specific angular momentum of the particle, respectively.

\begin{figure}[ht]
\begin{center}
\includegraphics[width=8cm]{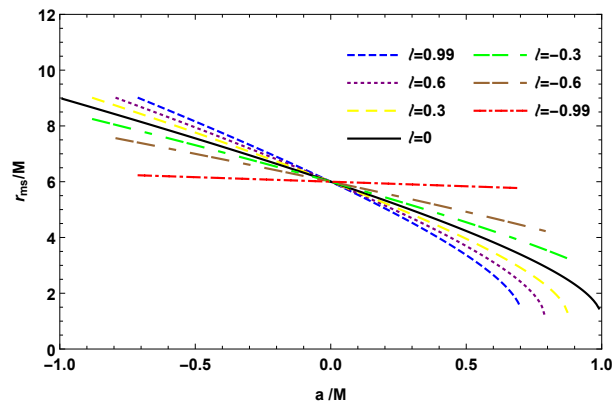}
\caption{The marginally stable orbit radius $r_{ms}$ with
the LV parameter $\it l$ for the thin disk around the
rotating Kerr-like black hole.}\label{frms}
\end{center}
\end{figure}

The circular equatorial orbits obey the conditions $V_{eff}=0$,
$V_{eff,r}=0$ and $V_{eff,\theta} =0$ \cite{CBa3,CBa4,JGa}. Due to
the refection symmetry of the metric (\ref{metric0}) with respect to
the equatorial plane, one can find that the condition
$V_{eff,\theta} =0$ is satisfied naturally for the particles
locating at the plane $\theta=\pi/2$. Making use of these
conditions, we can get the specific energy $\tilde{E}$, the specific
angular momentum $\tilde{L}$ and the angular velocity $\Omega$ of
the particle moving in circular orbit on the equatorial plane in the
rotating Kerr-like black hole spacetime
\begin{eqnarray}
&&\tilde{E}=-\frac{g_{tt}+g_{t\phi}\Omega}{\sqrt{-g_{tt}-2g_{t\phi}\Omega-g_{\phi\phi}\Omega^2}}=\frac{a \sqrt{(\it l+1) M}+\sqrt{r} (r-2 M)}{\sqrt{2 a r^{3/2} \sqrt{(\it l+1) M}-3 M r^2+r^3}},
\label{sE}\\
&&\tilde{L}=\frac{g_{t\phi}+g_{\phi\phi}\Omega}{\sqrt{-g_{tt}-2g_{t\phi}\Omega-g_{\phi\phi}\Omega^2}}=\frac{\sqrt{M} \left(a \left(a(1+\it l)-2 \sqrt{(1+\it l) M r}\right)+r^2\right)}{\sqrt{2 a r^{3/2} \sqrt{(1+\it l) M}-3 M r^2+r^3}}\;, \label{sL}\\
&&\Omega=\frac{d\phi}{dt}=\frac{-g_{t\phi,r}+\sqrt{g_{t\phi,r}^2-g_{tt,r}g_{\phi\phi,r}}}{g_{\phi\phi,r}}=\frac{1}{\frac{r^{3/2}}{\sqrt{M}}+a\sqrt{1+\it l}}.
\end{eqnarray}
The marginally stable orbit radius is determined by the condition
or $V_{eff}=0$, $V_{eff,r}=0$ and $V_{eff ,rr}=0$.
In fact , we can also use the conditions $d\tilde{E}/dr$ or  $d\tilde{L}/dr$ \cite{Page} to obtain it.
With the help of the equation (\ref{sE}), the marginally stable orbit on the equatorial plane is given by the following equations
\begin{eqnarray}
r(r-6 M)+8 a \sqrt{M r (1+\it l)}-3 a^2 (1+\it l)=0
\label{msob}\\
\end{eqnarray}
which give the radius of the marginally stable orbit as
\begin{eqnarray}
r_{ms}&=&M \left(3+B-\sqrt{(3-A)(3+A+2B}\right),\\\nonumber
A&=&\sqrt[3]{1-\frac{a^2 (l+1)}{M^2}} \left(\sqrt[3]{1-\frac{a \sqrt{l+1}}{M}}+\sqrt[3]{\frac{a \sqrt{l+1}}{M}+1}\right)+1,\\\nonumber
B&=&\sqrt{\frac{3 a^2 (l+1)}{M^2}+A^2}.
\label{isco}\\
\end{eqnarray}
For $a=0$, the marginally stable orbit radius $r_{ms}$ is always equals to $6M$ and independent of the LV parameter $\it l$. In Fig.(\ref{frms}), we plot the variety of the marginally stable orbit radius
$r_{ms}$ with the LV parameter $\it l$ in the rotating
Kerr-like black hole. It shows that the $r_{ms}$ decreases with $\it l$ in the prograde orbit$(a>0)$, and increases with $\it l$ in the retrograde orbit$(a<0)$.

\section{The properties of thin accretion disks in the
rotating Kerr-like black hole spacetime}

In this section we will adopt the steady-state thin accretion disk
model to study the accretion process in the thin disk around the
rotating Kerr-like black hole and probe how the LV parameter
$\it l$ affects the energy flux, the conversion efficiency, the
radiation temperature and the spectra of the disk in this
background.  In the steady-state accretion disk models, the
accreting matter in the disk can be described by an anisotropic
fluid with the energy-momentum tensor \cite{Page,Thorne}
\begin{eqnarray}
T^{\mu\nu}=\varepsilon_0
u^{\mu}u^{\nu}+2u^{(\mu}q^{\nu)}+t^{\mu\nu},
\end{eqnarray}
where the quantities $\varepsilon_0$, $q^{\mu}$ and $t^{\mu\nu}$
denotes the rest mass density, the energy flow vector and the stress
tensor of the accreting matter, respectively,  which are defined in
the averaged rest-frame of the orbiting particle with four-velocity
$u^{\mu}$. In the averaged rest-frame, we have $u_{\mu}q^{\mu}=0$
and $u_{\mu}t^{\mu\nu}=0$ since both $q^{\mu}$ and $t^{\mu\nu}$ is
orthogonal to $u^{\mu}$ \cite{Page,Thorne}.
\begin{figure}[ht]
\begin{center}
\includegraphics[width=6.4cm]{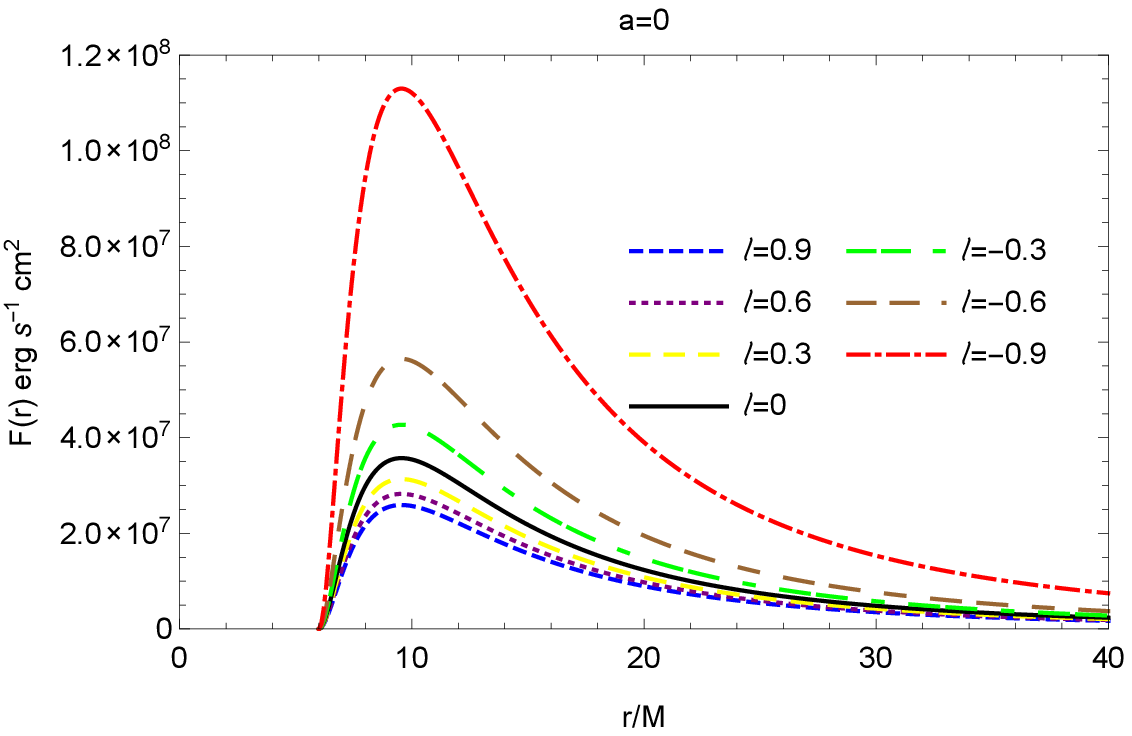}\;\;\includegraphics[width=6.4cm]{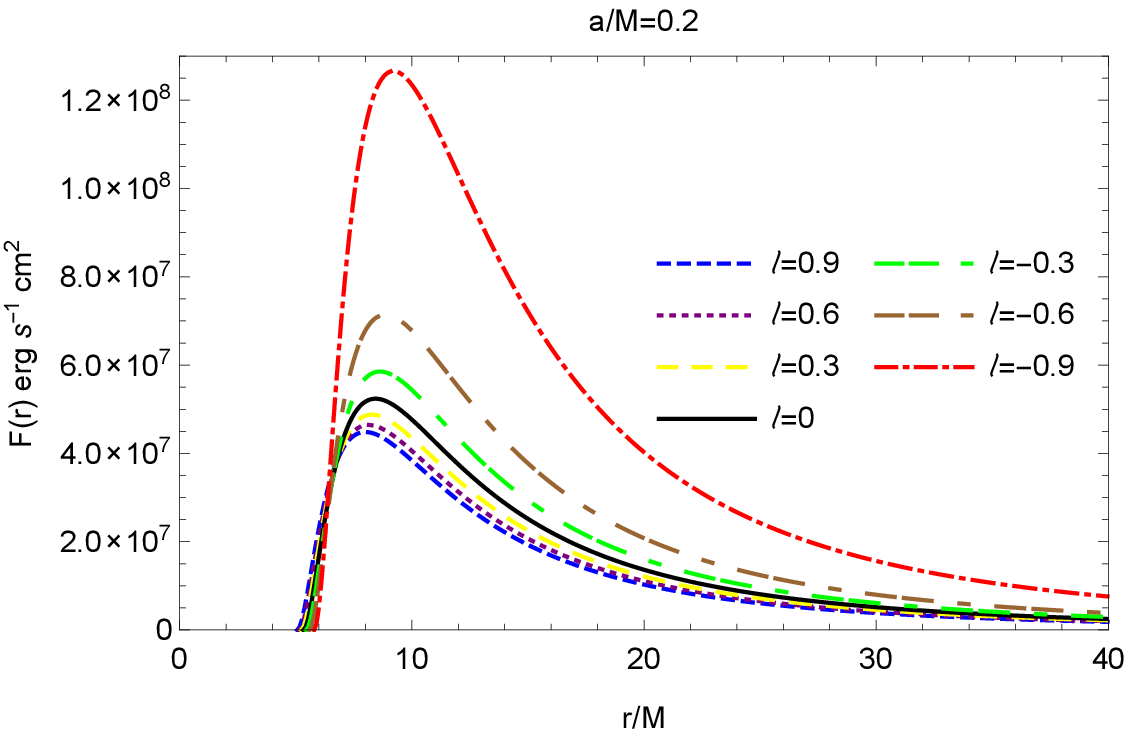}\;\;\\
\includegraphics[width=6.4cm]{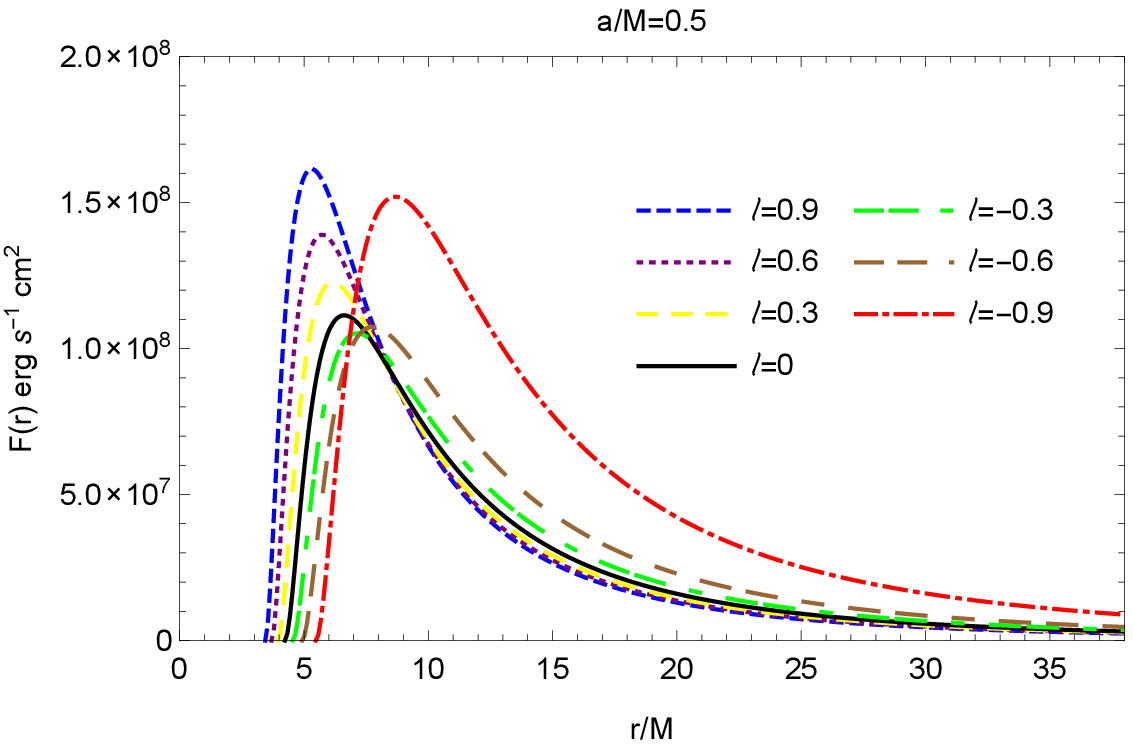}\;\;\includegraphics[width=6.4cm]{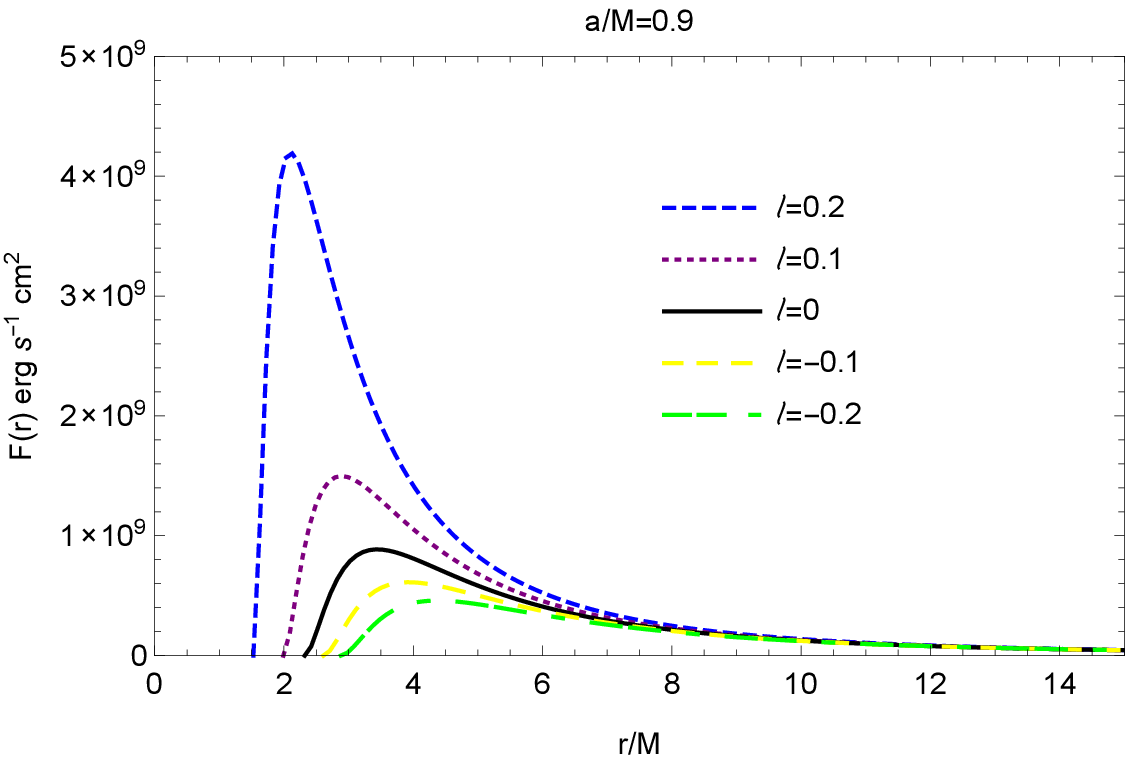}\;\;\caption{The energy flux $F(r)$ with the LV parameter $\it l$ in the
thin disk around the rotating Kerr-like black hole. Here, we set the total mass of the
black hole $M=10^6M_{\odot}$ and the mass accretion rate
$\dot{M_0}=10^{-12}M_{\odot}\;yr^{-1}$.}\label{flux}
\end{center}
\end{figure}

In background of the rotating Kerr-like black hole, one can find that
the time-averaged radial structure equations of the thin disk can be
expressed as
\begin{eqnarray}
&&\dot{M_0}=-2\pi\sqrt{-G}\Sigma(r) u^{r}=\text{Const},\\
&&\bigg[\dot{M_0}\tilde{E}- 2\pi\sqrt{-G}\Omega
W_{\phi}^{\;r}\bigg]_{,r}
=2\pi\sqrt{-G}F(r)\tilde{E},\label{Ws1}\\
&&\bigg[\dot{M_0}\tilde{L}-2\pi\sqrt{-G}W_{\phi}^{\;r}\bigg]_{,r}
=2\pi\sqrt{-G}F(r)\tilde{L},\label{Ws2}
\end{eqnarray}
with
\begin{eqnarray}
\Sigma(r)=\int^{H}_{-H}\langle\varepsilon_0\rangle
\;dz,\;\;\;\;\;\;\; W_{\phi}^{\;r}=\int^{H}_{-H}\langle
t_{\phi}^{\;r} \rangle
\;dz,\;\;\;\;\;\;\;\sqrt{-G}=\sqrt{1+\it l} r,
\end{eqnarray}
where $\Sigma(r)$ and  $W_{\phi}^{\;r}$ are the averaged rest mass
density and the averaged torque, respectively. The quantity $\langle
t_{\phi}^{\;r}\rangle$ is the average value of the $\phi-r$
component of the stress tensor over a characteristic time scale
$\Delta t$ and the azimuthal angle $\Delta\phi=2\pi$. With the
energy-angular momentum relation for circular geodesic orbits
$\tilde{E}_{,r}=\Omega \tilde{L}_{,r}$, one can eliminate
$W_{\phi}^{\;r}$ from Eqs.(\ref{Ws1}) and (\ref{Ws2}), and then
obtain the expression of the energy flux in the mass accretion
process
\begin{eqnarray}
F(r)=-\frac{\dot{M_0}}{4\pi\sqrt{-G}}
\frac{\Omega_{,r}}{(\tilde{E}-\Omega\tilde{L})^2}\int^{r}_{r_{ms}}
(\tilde{E}-\Omega\tilde{L})\tilde{L}_{,r}dr.\label{enf}
\end{eqnarray}

Here, we consider the mass accretion driven by black holes with a
total mass of $M=10^6M_{\odot}$, and with a mass accretion rate of
$\dot{M_0}=10^{-12}M_{\odot}\;yr^{-1}$ \cite{Harko}. In Fig. (\ref{flux}), we
plot the total energy flux $F(r)$ radiated by a thin disk around
the Kerr-like black hole for the LV parameter
$\it l$ and rotation parameter $a$. As particles is around the Schwarzschild-like black hole ($a=0$), the energy
flux $F(r)$ originates from the same value and then decreases with of the LV parameter $\it l$. However, the position of the peak value of $F(r)$ is at the same place of $r$. The main mathematical reason is that the marginally stable orbit radius $r_{ms}$, the specific energy $\tilde{E}$, the specific angular momentum $\tilde{L}$ and the angular velocity $\Omega$ of
the particle moving in circular orbit on the equatorial plane in the
Schwarzschild-like black hole have no changes compared with the Schwarzschild case.
As the rotation parameter $a$ increases from 0 to $M$. the increase of the energy flux $F(r)$  with the positive LV parameter $\it l$  is larger than that  decrease of the energy flux $F(r)$ with the negative LV parameter $\it l$. Thus, for the rapidly
rotating Kerr-like black hole, the energy flux $F(r)$ increases with  $\it l$.

\begin{figure}[ht]
\begin{center}
\includegraphics[width=7cm]{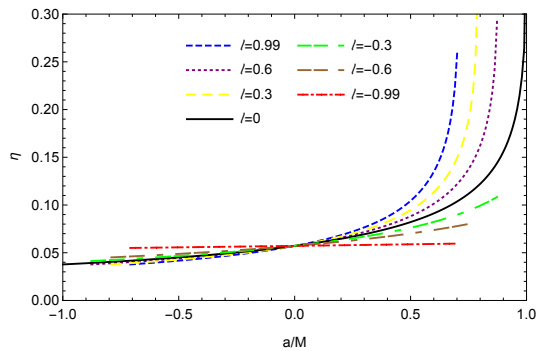}
\caption{The efficient $\eta$ with the parameter
$\it l$ for the thin disk around the rotating Kerr-like black
hole.}\label{feta}
\end{center}
\end{figure}

In the mass accretion process around a black hole, the conversion
efficiency is an important characteristic quantity which describes
the capability of the central object converting rest mass into
outgoing radiation. In general, the conversion efficiency can be
given by the ratio of two rates measured at infinity
\cite{sdk2,Page}: the rate of the radiation energy of photons
escaping from the disk surface to infinity and the mass-energy
transfer rate of the central compact object in the mass accretion.
If all the emitted photons can escape to infinity, one can find that
the efficiency $\eta$ is determined by the specific energy of a
particle at the marginally stable orbit $r_{ms}$\cite{Thorne}
\begin{eqnarray}
\eta=1-\tilde{E}_{ms}.\label{effi}
\end{eqnarray}
The dependence of the conversion efficiency $\eta$ on the LV
parameter $\it l$ is plotted in Fig.(\ref{feta}). It shows that the $\eta$ increases with $\it l$ in the prograde orbit$(a>0)$, and decreases with $\it l$ in the retrograde orbit$(a<0)$. This means that the
conversion efficiency of the thin accretion disk in the
Kerr-like black hole with the larger positive
$\it l$ is more higher than that the Kerr case.
\begin{figure}[ht]
\begin{center}
\includegraphics[width=6.4cm]{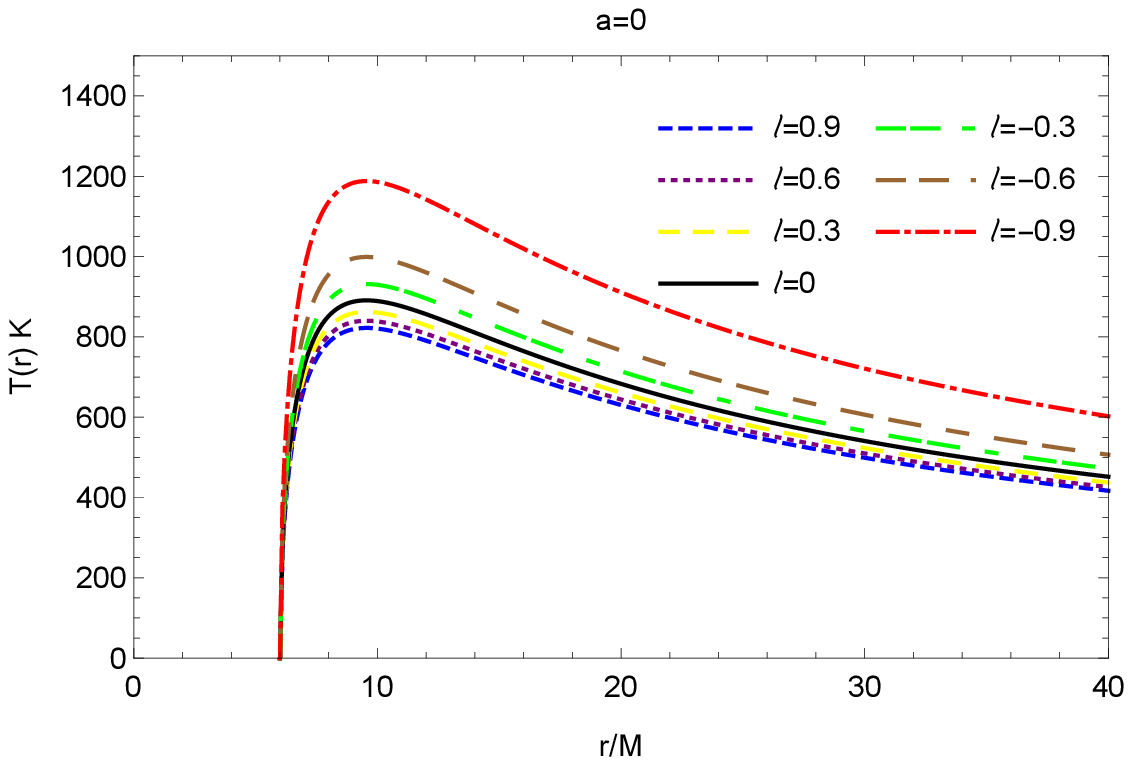}
\includegraphics[width=6.4cm]{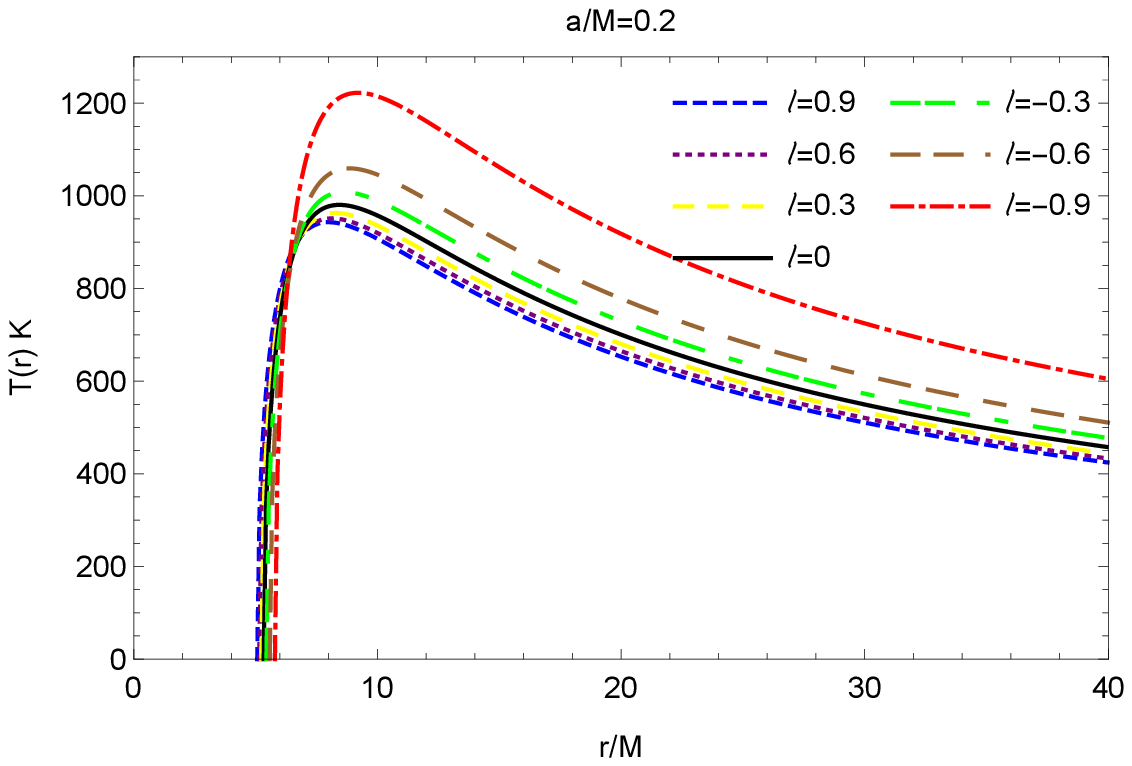}\\
\includegraphics[width=6.4cm]{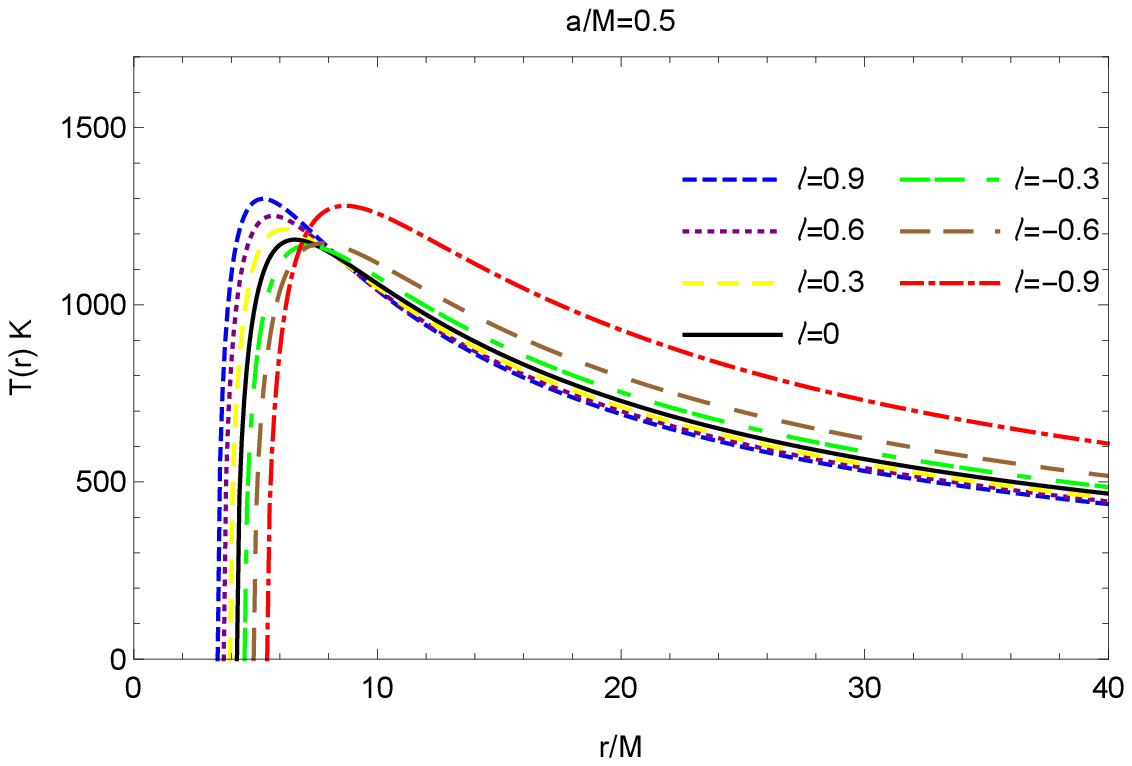}\includegraphics[width=6.4cm]{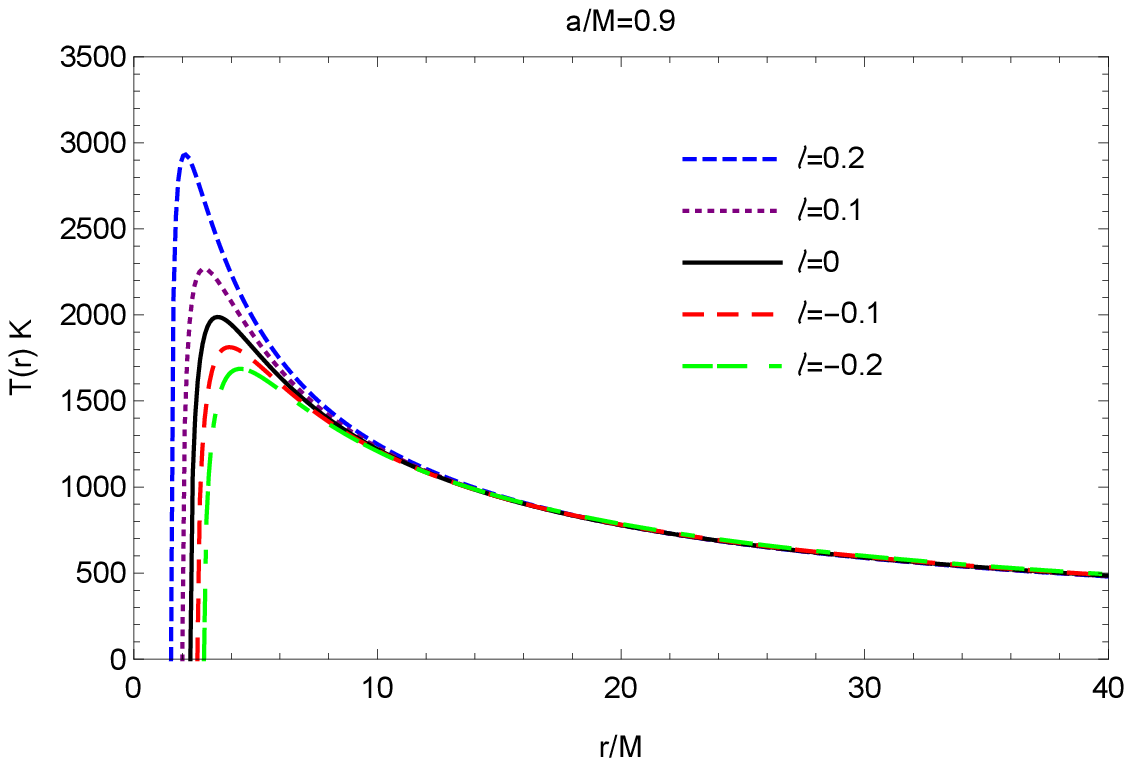}
\caption{The temperature $T$  with the LV parameter
$\it l$ in the thin disk around the rotating Kerr-like black hole. Here, we set the
total mass of the black hole $M=10^6M_{\odot}$ and the mass
accretion rate $\dot{M_0}=10^{-12}M_{\odot}\;yr^{-1}$. }\label{temper}
\end{center}
\end{figure}

Let us now to probe the effects of $it l$ on the radiation
temperature and the spectra of the disk around the rotating Kerr-like
black hole. In the steady-state thin disk model \cite{Page,Thorne},
it is assumed generally that the accreting matter is in
thermodynamic equilibrium, which means that the radiation emitted by
the disk surface can be considered as a perfect black body
radiation. The radiation temperature $T(r)$ of the disk is related
to the energy flux $F(r)$ through the expression
$T(r)=[F(r)/\sigma]^{1/4}$, where $\sigma$ is the Stefan-Boltzmann
constant. This means that the dependence of $T(r)$ on $\it l$ is
similar to that of the energy flux $F(r)$ on $\it l$, which is
also shown in Fig.(\ref{temper}).  Repeating the operations in \cite{Torres},
one can find that the observed luminosity $L(\nu)$ for the thin
accretion disk around the rotating Kerr-like black hole can be
expressed as
\begin{eqnarray}
L(\nu)=4\pi d^2I(\nu)=\frac{8\pi
h\cos{\gamma}}{c^2}\int^{r_f}_{r_{i}}\int^{2\pi}_{0}\frac{\nu_e^3\sqrt{-G}dr
d\phi}{e^{h\nu_e/KT(r)}-1},\label{emspe}
\end{eqnarray}
The emitted frequency is given by $\nu_e=\nu(1+z)$, where the
redshift factor can be written as
\begin{eqnarray}
1+z = \frac{1+\Omega r \sin\phi\sin\gamma}{\sqrt{-g_{tt}-2\Omega
g_{t\phi}-\Omega^2g_{\phi\phi}}},
\end{eqnarray}
where we have neglected the effect of light bending
\cite{CBa3,Luminet, Bhattacharyya}.
\begin{figure}[ht]
\begin{center}
\includegraphics[width=5.3cm]{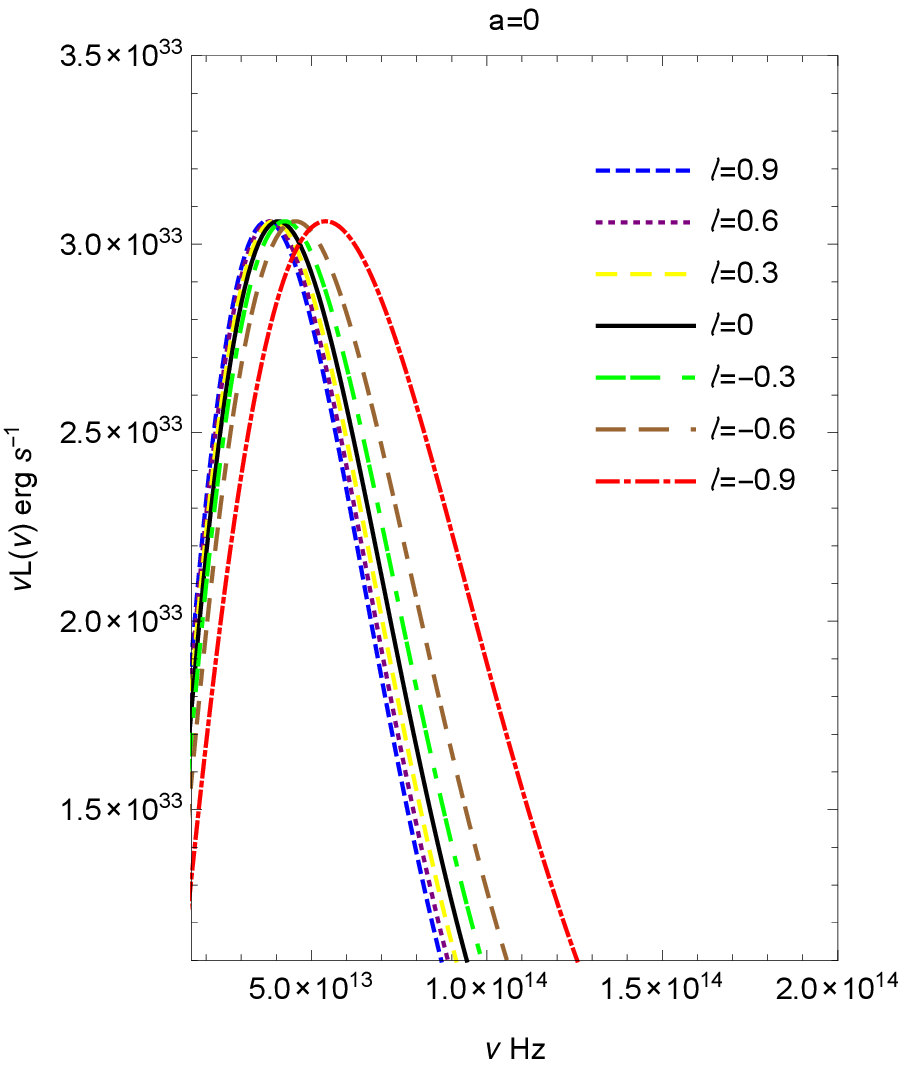}\includegraphics[width=6.2cm]{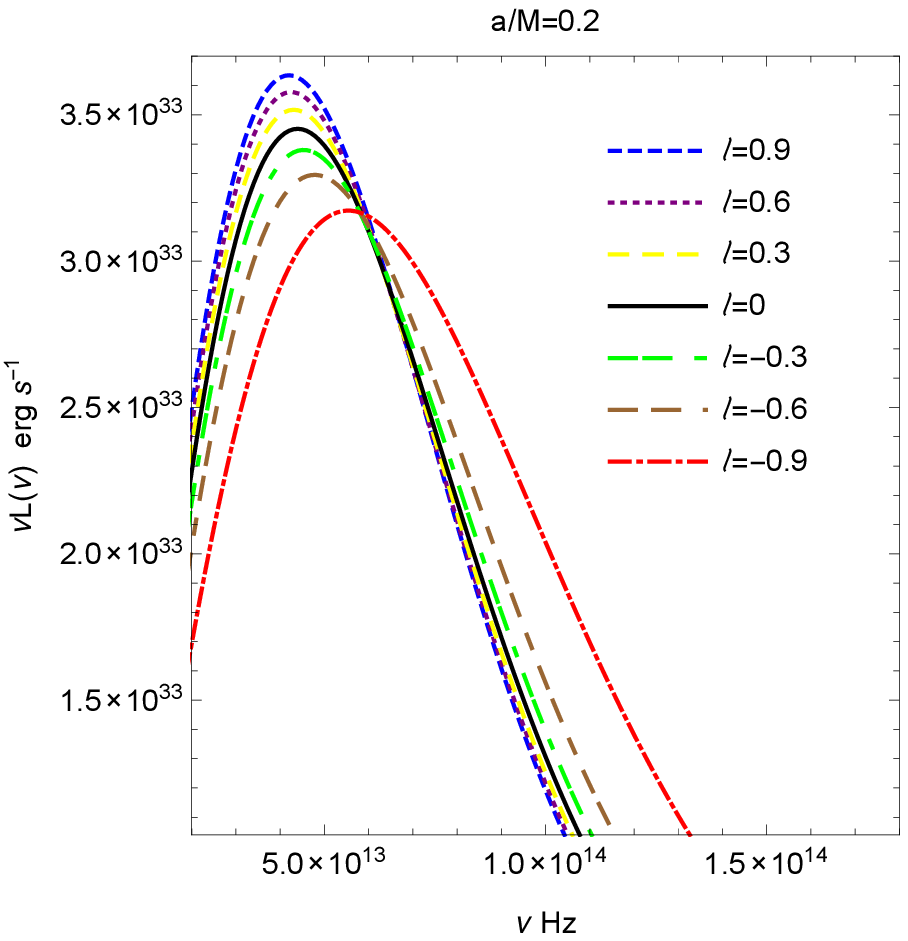}\\
\includegraphics[width=6.2cm]{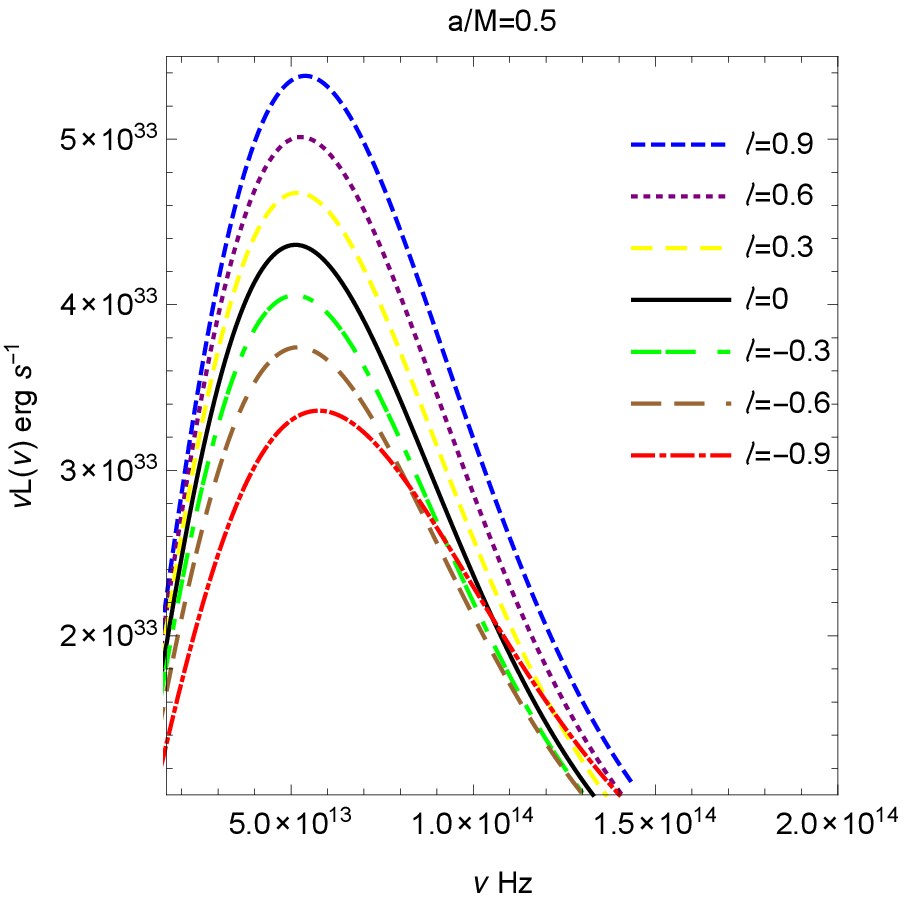}\includegraphics[width=5.5cm]{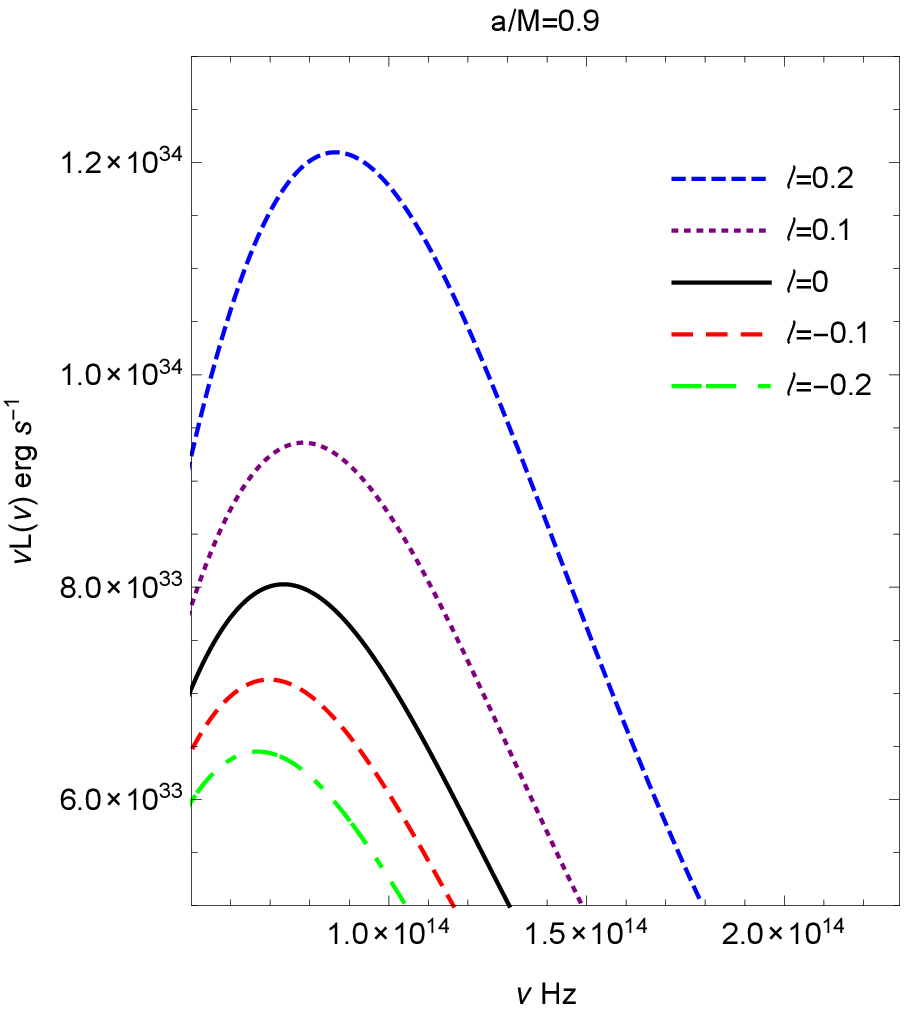}
\caption{ The emission spectrum with the $LV$
parameter $\it l$ in the thin disk around the rotating Kerr-like
black hole. Here,
we set the total mass of the black hole $M=10^6M_{\odot}$, the mass
accretion rate $\dot{M_0}=10^{-12}M_{\odot}\;yr^{-1}$ and the disk
inclination angle $\gamma=0^{\circ}$.}\label{lumin}
\end{center}
\end{figure}
The quantity $d$ is the distance to the source, $I(\nu)$ is the
thermal energy flux radiated by the disk, and  $\gamma$ is the disk
inclination angle. The quantities $r_f$ and $r_i$ are the outer and
inner border of the disk, respectively. In order to calculate the
luminosity $L(\nu)$ of the disk, we choose $r_i=r_{ms}$ and
$r_f\rightarrow \infty$ since the flux over the disk surface
vanishes at $r_f\rightarrow \infty$ in the rotating Kerr-like black
hole spacetime. Resorting to numerical method, we calculate the
integral (\ref{emspe}) and present the spectral energy distribution
of the disk radiation in Fig.(\ref{lumin}). As particles is around the Schwarzschild-like black hole ($a=0$), 
the smaller value of $\it l$ only leads to the higher cut-off frequencies, and the peak value of observed luminosity
has no changes and is independent of $\it l$. The main mathematical reason is the same as the case of energy flux $f(r)$.
As the rotation parameter $a$ increases from 0 to $M$. the increase of both the cut-off frequencies and observed luminosity  with the positive LV parameter $\it l$  is larger than that decrease of both the cut-off frequencies and observed luminosity with the negative LV parameter $\it l$. Thus, for the rapidly rotating Kerr-like black hole, both the cut-off frequencies and observed luminosity increase with  $\it l$.  Moreover, we also
find that the effect of the LV 
parameter $\it l$ on the spectra becomes more distinct for the rapidly rotating Kerr-like black hole.

\section{summary}

 In summary, we
have studied the properties of the thin accretion disk around the
rotating Kerr-like black hole in Einstein-bumblebee gravity model. we have
carried out an analysis of the marginally stable orbit radius $r_{ms}$.
 We have also presented the conversion
efficiency $\eta$ of the accreting mass into radiation, and we
have showed that the Kerr-like black hole with the positive parameter $\it l$ are much more efficient in converting the
accreting mass into radiation than their Kerr black holes
counterparts. Our results show that the
LV parameter $\it l$ imprints in the energy flux,
temperature distribution and emission spectra of the disk. For the Schwarzschild-like black hole ($a=0$),
 the larger LV parameter $\it l$ diminishes the energy flux, the radiation temperature, and
cut-off frequency of the thin accretion disk. the peak value of both the observed luminosity and conversion
efficiency is independent of $\it l$. The main mathematical reason is that the marginally stable orbit radius $r_{ms}$, the specific energy $\tilde{E}$, the specific angular momentum $\tilde{L}$ and the angular velocity $\Omega$ of
the particle moving in circular orbit on the equatorial plane in the
Schwarzschild-like black hole have no changes compared with the Schwarzschild case.
As the rotation parameter $a$ increases from $0$ to $0.9M$, the larger(smaller) LV parameter $\it l$ increases(diminishes) more distinct the energy flux, the radiation temperature, the observed luminosity and
cut-off frequency of the thin accretion disk.  Since the energy flux, the temperature distribution
of the disk, the spectrum of the emitted black body radiation, as well as the conversion efficiency show, Kerr-like black hole in Einstein-bumblebee gravity model, significant differences as compared to the general relativistic
case, the determination of these observational quantities
could discriminate, at least in principle, between standard general relativity and Einstein-bumblebee gravity.

\section{\bf Acknowledgments}

Changqing's work was supported by NNSFC No.11447168, the Natural Science
Foundation of of Hunan University of
Humanities Science and Technology No. 2016PY06, the State Scholarship Fund of
China No. 201708430087. Chikun's work was
supported by the National Natural Science Foundation of China under
Grant Nos11247013; Hunan Provincial Natural Science Foundation of
China under Grant Nos. 12JJ4007 and 2015JJ2085.  J. Jing's
work was partially supported by the National Natural Science
Foundation of China under Grant No.10935013; 973 Program Grant No.
2010CB833004.

\end{document}